\documentclass[twocolumn,preprintnumbers,amsmath,amssymb]{revtex4}   %PLB format%
\usepackage{epsfig}
\usepackage{dcolumn}   % needed for some tables
\usepackage{bm}        % for math
\usepackage{amssymb}
\begin{document}

\title{Particle Ratios as a Probe of the QCD Critical Temperature}

\author{J. Noronha-Hostler$^{1}$$^{2}$$^{3}$}
\author{H. Ahmad $^{2}$}
\author{J. Noronha $^{4}$}
\author{C. Greiner$^{2}$}
\affiliation{$^{1}$Frankfurt Institute for Advanced Studies (FIAS), Frankfurt am Main, Germany}
\affiliation{$^{2}$Institut f\"ur Theoretische Physik, Johann Wolfgang
Goethe--Universit\"at, Frankfurt am Main, Germany}
\affiliation{$^{3}$Helmholtz Research School, Frankfurt am Main, Germany}
\affiliation{$^{4}$Department of Physics, Columbia University, 538 West 120$^{th}$ Street, New York,
NY 10027, USA}

\begin{abstract}
We show how the measured particle ratios can be used to provide non-trivial information about the critical temperature of the QCD phase transition. This is obtained by including the effects of highly massive Hagedorn resonances on statistical models, which are used to describe hadronic yields. The inclusion of Hagedorn states creates a dependence of the thermal fits on the Hagedorn temperature, $T_H$, which is assumed to be equal to $T_c$, and leads to an overall improvement of thermal fits. We find that for Au+Au collisions at RHIC at $\sqrt{s_{NN}}=200$ GeV the best square fit measure, $\chi^2$, occurs at $T_c \sim 176$ MeV and produces a chemical freeze-out temperature of $172.6$ MeV and a baryon chemical potential of $39.7$ MeV.
\end{abstract}

\date{\today}
%\pacs{}
\maketitle

\section{Introduction}\label{intro}
Lattice QCD is the main non-perturbative theoretical tool used to probe bulk thermodynamics quantities of QCD such as its pressure, entropy density, and the speed of sound. The QCD phase transition at vanishing baryonic chemical potential is a (rapid) crossover where the thermodynamic quantities vary significantly near a critical temperature, whose value lies between $170-200$ MeV. In fact, according to the Bielefeld-BNL/RIKEN-Columbia collaboration (RBC-Bielefeld) the critical temperature is around $T_c=196$ MeV \cite{Cheng:2007jq} (although recently it has been concluded that the range could be $T_c=180-200$ MeV \cite{Bazavov:2009zn}) whereas the Budapest/Marseille/Wuppertal (BMW) collaboration has found a lower value $T_c=176$ MeV \cite{zodor}. Because the value of the critical temperature is vital to many phenomenological models of QCD, it is clearly important to find experimentally driven signals able to distinguish between these two critical temperature regions. We shall show in this Letter that thermal fits for the measured particle ratios in Au+Au collisions at $\sqrt{s_{NN}}=200$ GeV at RHIC can be used to determine the critical temperature of the QCD phase transition at nonzero baryonic chemical potential as long as effects from highly massive Hagedorn resonances are included.

Thermal fits computed within statistical models are normally used to reproduce hadron yield ratios in heavy ion collisions \cite{thermalmodels,StatModel,Schenke:2003mj,RHIC,Andronic:2005yp,Manninen:2008mg}.  Thermal models computed at AGS, SIS, SPS, and RHIC energies can be used to construct a chemical freeze-out line in the QCD phase diagram \cite{freezeoutline}.  For Au+Au collisions at RHIC$\;$ at $\sqrt{s_{NN}}=200$ GeV, specifically, estimates for the chemical freeze-out temperature and baryon chemical potential range from $T_{ch}=155-169$ MeV and $\mu_b=20-30$ MeV \cite{Andronic:2005yp,Manninen:2008mg,RHIC}.

Originally, it was thought that the chemical freeze-out temperature and the critical temperature coincided.  However, recent lattice results indicate a higher critical temperature, which leads to a difference of $\Delta T=7-45$ MeV between $T_c$ and $T_{ch}$.  At SPS this difference was explained by allowing hadrons, especially exotic anti-baryonic states, to be ``born" out of chemical equilibrium at $T_c$ and reach chemical equilibrium through multi-mesonic collisions \cite{Rapp:2000gy,Greiner} because chemical equilibration times of binary collisions are too long \cite{Koch:1986ud}. At RHIC, multi-mesonic collisions are no longer adequate to explain chemical equilibration times \cite{Kapusta,Huovinen:2003sa} and this has led some to believe that hadrons are ``born" in chemical equilibrium \cite{Stock:1999hm,Heinz:2006ur}.  A way out of this scenario involving an overpopulation of pions and kaons has been suggested in \cite{BSW}. Another solution that has provided very promising results is the inclusion of Hagedorn states, which are heavy resonances with an exponentially growing mass spectrum \cite{Hagedorn:1968jf}  that open up the phase space and help drive hadrons quickly into chemical equilibrium \cite{Greiner:2004vm,NoronhaHostler:2007jf,NoronhaHostler:2007fg,longpaper}.
When a reaction of the form $n\pi\leftrightarrow HS\leftrightarrow n\pi +X\bar{X}$ is used where $X\bar{X}=p\bar{p},\;K\bar{K},$ or $\Lambda\bar{\Lambda}$, hadrons are able to reach chemical equilibrium at about $T_{ch}\approx160$ MeV \cite{NoronhaHostler:2007jf,longpaper} using various lattice critical temperatures \cite{longpaper}.  Moreover, it was shown that the $K/\pi$ and $(B+\bar{B})/\pi$ ratios, where $B=p+n$, match RHIC data well \cite{NoronhaHostler:2007jf}.

Not only have Hagedorn states provided a mechanism for explaining the temperature difference between $T_c$ and $T_{ch}$, but they have also been used to find a low $\eta/s$ in the hadron gas phase \cite{NoronhaHostler:2008ju}, which nears the string theory bound $\eta/s=1/(4\pi)$ \cite{KSS}. Calculations of the trace anomaly including Hagedorn states also fits recent lattice results well and correctly describe the minimum of  $c_s^2$ near the phase transition found on the lattice \cite{NoronhaHostler:2008ju}. Furthermore, estimates for the bulk viscosity including Hagedorn states in the hadron gas phase indicate that $\zeta/s$ increases near $T_c$, which agrees with the general analysis done in \cite{Kharzeev:2007wb}.

Although Hagedorn's idea of an exponentially growing mass spectrum originated in the late 1960's, recent experimental results maintain an exponential mass spectrum albeit with a higher $T_c$ \cite{Broniowski:2004yh}. Moreover, it has been recently shown in \cite{Cohen:2009wq} that a Hagedorn spectrum does appear in QCD with a large number of colors. Moreover, thoughts on observing Hagedorn states in experiments are given in \cite{Bugaev:2008nu} and their usage as a thermostat in \cite{Moretto:2006zz}. A possible method for describing the cross sections of Hagedorn states was derived in \cite{Pal:2005rb}.

Since Hagedorn states have been shown to affect the chemical equilibration times, thermodynamic properties, and transport coefficients of hadron resonance gases close to $T_c$ it is natural to expect that they may also be relevant in the thermal description of particle ratios. Moreover, because Hagedorn states are dependent on the limiting Hagedorn temperature $T_H=T_c$, a relationship between the chemical freeze-out temperature and the critical temperature can be found by including Hagedorn states in thermal fits. This uniquely gives us the ability to distinguish between different critical temperature regions depending on the quality of the fit obtained using the statistical model.

%In this paper we use a statistical model that includes Hagedorn resonances in order to predict hadron yield ratios at RHIC from which we can fit to obtain $T_{ch}$ and $\mu_b$. We do not include any strangeness suppression factor such as $\gamma_s$ or in other words $\gamma_s=1$.  Our results indicate that not only do Hagedorn states improve the thermal fits over models that only include the resonances from the particle data group but they also indicate a lower critical temperature.

\section{Model}

In this paper we use a grand-canonical model to describe the particle densities from which we can calculate the corresponding ratios as described in detail in \cite{StatModel}.  We do not include any strangeness suppression factor or, in other words, we assume $\gamma_s=1$. In order to calculate the baryonic chemical potential $\mu_b$ and the strange chemical potential $\mu_s$ we use the following conservation relation
\begin{eqnarray}\label{eqn:cons}
0&=&\frac{\sum_i n_i S_i}{\sum_i n_i B_i},
\end{eqnarray}
which means that the total strangeness per baryon number is held at zero. There $n_i$ is the density of the $i^{th}$ particle that has a corresponding baryon number $B_i$ and strangeness $S_i$.

Hagedorn states are included in our hadron resonance gas model via the exponentially increasing density of states
\begin{equation}\label{eqn:fitrho}
    \rho(M)=\int_{M_{0}}^{M}\frac{A}{\left[m^2 +m_{r}^2\right]^{\frac{5}{4}}}e^{\frac{m}{T_{H}}}dm,
\end{equation}
which follows from Hagedorn's original idea that you have an exponentially growing mass spectrum that has a limiting temperature, $T_H$. Close to  $T_H$ the Hagedorn states become increasingly relevant and heavier resonances are ``formed" the closer you get to $T_H$.  We use the particles from the particle date group up until $M_0=2$ GeV and then we use Hagedorn states above 2 GeV. Additionally, A is the Hagedorn state ''degeneracy", M is the maximum mass, and $m_r=500$ MeV, which is taken from \cite{Hagedorn:1968jf,Broniowski:2004yh}.

In this paper we use two different scenarios regarding $T_H$.  First we assume that $T_H=T_c$, and then we consider the two different different lattice results for $T_c$: $T_c=196$ MeV \cite{Cheng:2007jq} and $T_c=176$ MeV \cite{zodor}. Futhermore, we take into account effects from repulsive interactions between the hadrons \cite{Kapusta:1982qd,Rischke:1991ke} via the following excluded-volume corrections \cite{Kapusta:1982qd}:
\begin{eqnarray}\label{eqn:cor}
T&=&\frac{T^*}{1-\frac{p_{pt}\left(T^*,\;\mu_b^*\right)}{4B}}\nonumber\\
\mu_b&=&\frac{\mu_b^*}{1-\frac{p_{pt}\left(T^*,\;\mu_b^*\right)}{4B}}\nonumber\\
p_{xv}(T,\mu_b)&=&\frac{p_{pt}\left(T^*,\mu_b^*\right)}{1-\frac{p_{pt}\left(T^*,\;\mu_b^*\right)}{4B}}\nonumber\\
\varepsilon_{xv}(T,\mu_b)&=&\frac{\varepsilon_{pt}\left(T^*,\mu_b^*\right)}{1+\frac{\varepsilon_{pt}\left(T^*,\;\mu_b^*\right)}{4B}}\nonumber\\
n_{xv}(T,\mu_b)&=&\frac{n_{pt}\left(T^*,\mu_b^*\right)}{1+\frac{\varepsilon_{pt}\left(T^*,\;\mu_b^*\right)}{4B}}\,.
\end{eqnarray}
Note that the system's temperature $T$, baryonic chemical potential $\mu_b$ as well as the thermodynamic functions (after volume corrections) are defined in terms of the quantities computed in the point particle (subscript pt) approximation (i.e., no volume corrections). Note that $B$ is equivalent to an effective MIT bag constant and is taken as a parameter in our model.

In order to find the maximum $M$ Hagedorn state masses and the degeneracy $A$, we fit our model to the thermodynamic properties of the lattice at zero chemical potential $\mu_b=0$.  In the RBC-Bielefeld collaboration the thermodynamical properties are derived from $\varepsilon-3p$, which is what we fit in order to obtain the parameters for the Hagedorn states.  In this case we set $T_H=196$ MeV and $A=0.5 \,{\rm GeV}^{3/2}$ and obtain $M=20$ GeV and $B=\left(340 \,{\rm MeV}\right)^4$. For the BMW collaboration the energy density is fitted and we fix $T_H=176$ MeV and $A=0.5\, {\rm GeV}^{3/2}$ and obtain $M=15$ GeV and $B=\left(250 \,{\rm MeV}\right)^4$. Additionally, a discretized version of the resonance spectrum is considered, which is separated into mass bins of 100 MeV. Only mesonic, non-strange Hagedorn states are considered in our calculations.

In our model we do not just consider the direct number of hadrons but also the indirect number that comes from other resonances.  For example, for pions we consider also the contribution from resonances such as $\rho$'s, $\omega$'s etc.  The number of indirect hadrons can be calculated from the branching ratios for each individual species in the particle data book \cite{Eidelman:2004wy}. Moreover, there is also a contribution from the Hagedorn states to the total number of pions, kaons, and so on as described in \cite{NoronhaHostler:2007jf,longpaper}.  Thus the total number of ``effective" pions can be described by
\begin{eqnarray}\label{eqn:effpi}
\tilde{N}_{\pi}&=&N_{\pi}+\sum_{i}N_{i}\langle n_{i}\rangle
\end{eqnarray}
whereas the total number of ``effective" protons, kaons, or lambdas (generalized as $X$) can be described by
\begin{eqnarray}\label{eqn:effbbkk}
\tilde{N}_{X}&=&N_{X}+\sum_{i}N_{i}\langle X_{i}\rangle
\end{eqnarray}
where $\langle X\rangle$ is the average number of $X=$ p's, K's, or $\Lambda$'s.
Here $N$ is the total number of each species and $\langle n_{i}\rangle $ is the average number of pions that each Hagedorn state decays into.

To determine $\langle X\rangle$ we use the multiplicities in Fig.\ 2 of Ref.\ \cite{Greiner:2004vm} from the microcanonical model in \cite{Liu} such that
\begin{eqnarray}\label{eqn:gamfit}
  p&=& 0.058\;m_{i}-0.10 \nonumber\\
   K^{+}&=&0.075\;m_{i}+0.047\nonumber\\
  \Lambda&=&0.04\;m_{i}-0.07.
\end{eqnarray}
Clearly, they are all dependent on the mass of the $i^{th}$ Hagedorn state.  Of course, in principle, the branching ratios of potential Hagedorn states are not known. Future measurements of high exotic hadronic resonances can be used to obtain these ratios in the future.  Following the principle of (maximum) missing information, we assume here that the branching ratios can be obtained from a microcanonical calculation. Such a description is, for instance, also appropriate for describing the annihilation of p and anti-p.

In order to get an idea of the quality of the thermal fits, we define $\chi^2$ as
\begin{equation}
\chi^2=\sum_i \frac{\left(R_i^{exp}-R_i^{therm}\right)^2}{\sigma_i^2}
\end{equation}
where $R_i^{therm}$ is our ratio of hadron yields calculated within our thermal model whereas $R_i^{exp}$ is the experimentally measured value of the hadron yield with its corresponding error $\sigma_i^2$.  In this paper we look at only the experimental values at mid-rapidity and we used only the systematic error given by each respective experiment. We vary the temperature and $\mu_b$ according to the conservation laws in Eq.\ (\ref{eqn:cons}), in order to get the smallest $\chi^2$.  We use the experimental data from both STAR \cite{STAR} and PHENIX \cite{PHENIX} for Au+Au collisions at RHIC at $\sqrt{s_{NN}}$=200 GeV.  Specifically, we observe the ratios: $\pi^{-}/\pi^{+}$, $\bar{p}/p$, $K^-/K^+$, $K^+/\pi^+$, $p/\pi^+$, and $(\Lambda+\bar{\Lambda})/\pi^+$.  All of which are calculated by STAR \cite{STAR}.  However, only $\pi^{-}/\pi^{+}$, $\bar{p}/p$, $K^-/K^+$, $K^+/\pi^+$, $p/\pi^+$ are given by PHENIX.  Because there is such a difference between $p/\pi^+$ from PHENIX and STAR we choose only the value from STAR so that we can compare are results to \cite{Andronic:2005yp} where they also exclude $p/\pi^+$ from PHENIX.  It should be noted that Ref.\ \cite{Andronic:2005yp} includes more ratios than we do such as multi-strange particles and resonances, which are not included in this paper.  This is because the purpose of this paper is not to confirm their results, which have already been confirmed in \cite{Manninen:2008mg}, but rather to compare thermal fits that include the contribution of Hagedorn states and those that exclude them.

\section{Results}

\begin{figure}[h]
\centering
\epsfig{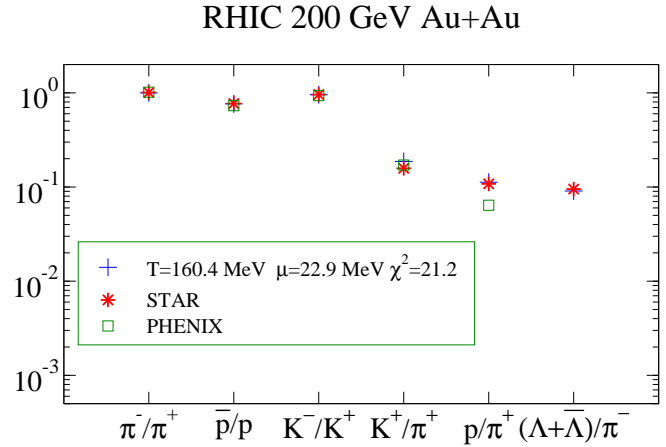}
\caption{Thermal fits for Au+Au collisions at RHIC at $\sqrt{s_{NN}}=200$ GeV when no Hagedorn states are present.  }
\label{fig:noHS}
\end{figure}

The following results are given for the minimal $\chi^2$ for a given $\mu_b$ and $T_{ch}$.  Initially, we found the thermal fit for a hadron gas excluding Hagedorn states, which is shown in Fig.\ \ref{fig:noHS}. There $T_{ch}=160.4$ MeV, and $\mu_b=22.9$ MeV, which gave $\chi^2=21.2$.  Our resulting temperature and baryonic chemical potential are almost identical to that in \cite{Andronic:2005yp} where $T_{ch}=160.5$ and $\mu_b=20$ MeV.

\begin{figure}[h]
\centering
\epsfig{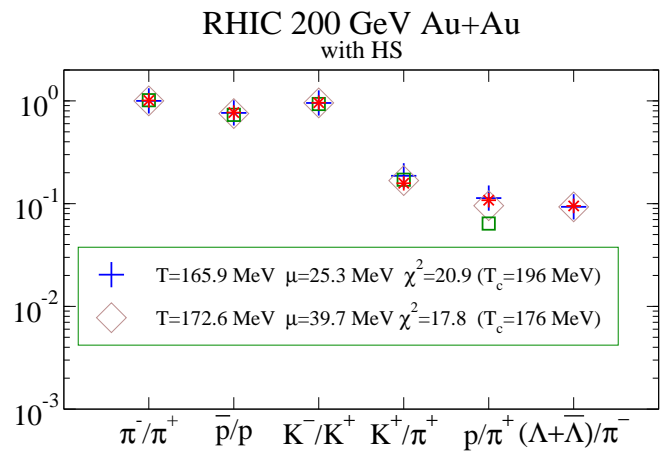}
\caption{Thermal fits including Hagedorn states for Au+Au collisions at RHIC at $\sqrt{s_{NN}}=200$ GeV.  }
\label{fig:HS}
\end{figure}

The inclusion of Hagedorn states is our primary interest.  Starting with the fit for the RBC-Bielefeld collaboration, we obtain $T_{ch}=165.9$ MeV, $\mu_b=25.3$ MeV, and $\chi^2=20.9$, which is shown in Fig.\ \ref{fig:HS}.  The $\chi^2$ is actually slightly smaller than in Fig.\ \ref{fig:noHS}.  The contributions of the Hagedorn states to the total number of the various species at this temperature and chemical potential are shown in Tab.\ \ref{tab:HScon}.

When we consider the lattice results from BMW, which are at the lower end of the critical temperature spectrum where $T_c=176$ MeV, we find $T_{ch}=172.6$ MeV, $\mu_b=39.7$ MeV, and $\chi^2=17.8$. The lower critical temperature seems to have a significant impact on the thermal fit. The lower $\chi^2$ is due to the larger contribution of Hagedorn states at at $T_{ch}=172.6$ MeV, which is much closer to $T_c$.  The contributions of the Hagedorn states to the total number of the various species at this temperature and chemical potential are about $30-50\%$ as shown in Tab.\ \ref{tab:HScon}.

\begin{figure}[h]
\centering
\epsfig{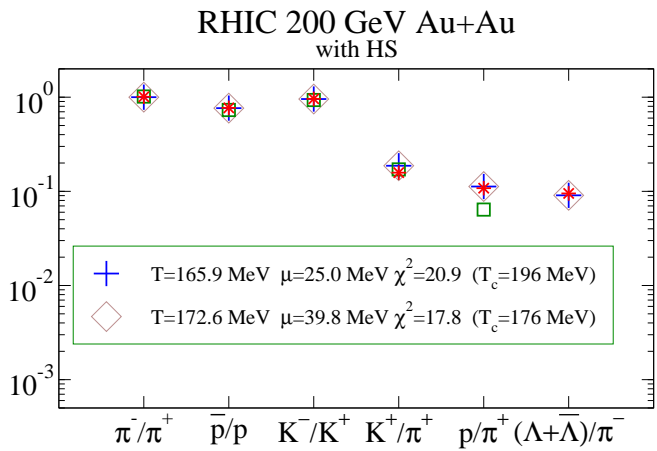}
\caption{Thermal fits including Hagedorn states for Au+Au collisions at RHIC at $\sqrt{s_{NN}}=200$ GeV when the maximum mass of the Hagedorn states is doubled.}
\label{fig:M2}
\end{figure}

The difference in the $\chi^2$'s for BMW and RBC-Bielefeld collaboration is directly related to the contribution of Hagedorn states in the model.  Because the RBC-Bielefeld critical temperature region is significantly higher than its corresponding chemical freeze-out temperature the contribution of the Hagedorn states is minimal at only 4-11$\%$ (see Tab.\ \ref{tab:HScon}.).  To further prove this point we can vary the parameters that define the influence of Hagedorn states in the model.  If, for instance, we double the maximum mass we see in Fig.\ \ref{fig:M2} that our thermal fits are not affected.  This effect arises the true limiting temperature after volume corrections is larger than the critical temperature \cite{NoronhaHostler:2008ju}. The effects of changing the maximum mass are only seen at temperatures larger than the critical temperature, which are not considered in this study.

\begin{figure}[h]
\centering
\epsfig{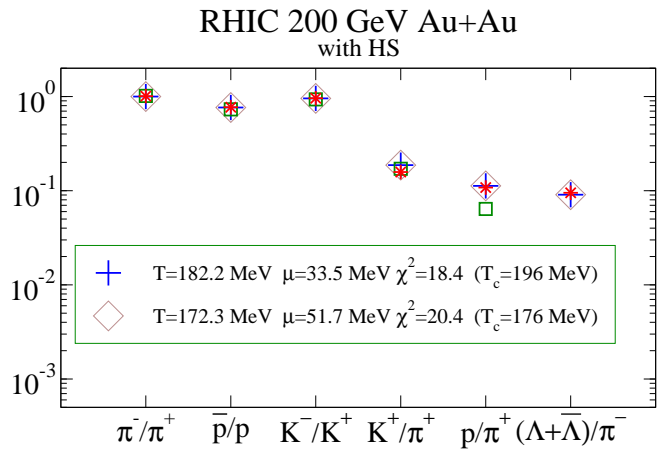}
\caption{Thermal fits including Hagedorn states for Au+Au collisions at RHIC at $\sqrt{s_{NN}}=200$ GeV when the  degeneracy of the Hagedorn states is doubled.}
\label{fig:A2}
\end{figure}

\begin{table}
\begin{center}
 \begin{tabular}{|c|c|c|c|c|c|c|}
 \hline
 $T_c$ (MeV) & $A$ (GeV$^{3/2}$) & $M$ (GeV) & $\pi$'s & $K$'s & $p$'s & $\Lambda$'s\\
 \hline
 176 & 0.5 & 15 & $48.5\%$  & $41.6\%$ &  $29.1\%$ & $41.0\%$ \\
 196 & 0.5 & 20 & $11.2\%$ & $10.5\%$ & $4.7\%$ & $6.2\%$ \\
  176 & 0.5 & 30 & $48.7\%$  & $41.6\%$ &  $29.1\%$ & $41.2\%$ \\
 196 & 0.5 & 40 & $11.2\%$ & $10.5\%$ & $4.7\%$ & $6.2\%$ \\
  176 & 1.0 & 15 & $62.5\%$  & $56.0\%$ &  $40.6\%$ & $53.4\%$ \\
 196 & 1.0 & 20 & $44.0\%$  & $38.9\%$  & $21.9\%$ & $30.3\%$\\
 \hline
 \end{tabular}
 \end{center}
 \caption{Contribution of the Hagedorn states to the total number of hadron species.}\label{tab:HScon}
 \end{table}

\begin{table}
\begin{center}
 \begin{tabular}{|c|c|c|c|c|c|}
 \hline
 $T_c$ (MeV) & $A$ (GeV$^{3/2}$) & $M$ (GeV) & $T_c$ (MeV) & $\mu_b$ (MeV) & $\chi^2$ \\
 \hline
 176 & 0.5 & 15 & 172.6 & 39.7 & 17.8\\
 196 & 0.5 & 20 & 165.9  & 25.3 & 20.9 \\
  176 & 0.5 & 30 & 172.6 & 39.8 & 17.8  \\
 196 & 0.5 & 40 & 165.9  & 25.0 & 20.9  \\
  176 & 1.0 & 15 & 172.3  & 51.7 &  20.4  \\
 196 & 1.0 & 20 & 182.2 & 33.5  & 18.4\\
 \hline
 \end{tabular}
 \end{center}
 \caption{Comparison of the chemical freeze-out temperature, baryonic chemical potential, and $\chi^2$ for various fits including Hagedorn states.}\label{tab:all}
 \end{table}

While the maximum mass does not affect the quality of the fit, the parameter $A$, which is essentially the degeneracy of the Hagedorn states, does. The results of this are shown in Fig.\  \ref{fig:A2}.  If we double $A$ then we find that the minimum $\chi^2$ for $T_c=196$ MeV has dropped down to $\chi^2=18.4$, which is only slightly higher than the best fit for $T_c=176$ MeV in Fig.\  \ref{fig:HS}.  This indicates that at $T_c=196$ MeV more Hagedorn states would be needed in order to get a better fit. However, we also see that for $T_c=176$ MeV and A=1.0 GeV$^{3/2}$ that $\chi^2=20.4$.  The reason for this is that there is an overpopulation of Hagedorn states.  If we look at the contribution of Hagedorn states to the individual particle species we see that the optimal contribution of Hagedorn states is around $\approx40\pm10\%$, which is what we get for the fits $T_c=176$ MeV, A=0.5 GeV$^{3/2}$, M=15 GeV and  $T_c=196$ MeV, A=1.0 GeV$^{3/2}$, M=20 GeV as seen in Tab.\ \ref{tab:HScon}.

A summary of our results is seen in Tab.\ \ref{tab:all}.  We find that the inclusion of Hagedorn states should not only provide a better fit but they also affect the chemical freeze-out temperature and the baryonic chemical potential. The more mesonic Hagedorn states are present the larger $\mu_b$ becomes.  Furthermore, our fits also have higher $T_{ch}$'s than seen in the fit without the effects of Hagedorn states.

\section{Conclusions}

In this paper we assumed that the particle ratios measured in Au+Au collisions at RHIC at $\sqrt{s_{NN}}=200$ GeV admit a purely statistical description at chemical freeze-out. Our results for thermal fits without Hagedorn states concur well with other thermal fit models \cite{Andronic:2005yp} where the chemical freeze-out temperature ($T_{ch}=160.4$ MeV)  is almost identical and the baryonic chemical potential ($\mu_b=22.9$ MeV)  is only slightly larger. The thermal fit with the known particles in the particle data group provides a decent fit with $\chi^2=21.2$.  However, the inclusion of Hagedorn states provides an even better fit to the experimental data. In fact, we find $\chi^2=17.8$, $T_{ch}=172.6$ MeV, and $\mu_b=39.7$ MeV for the BMW collaboration while for the RBC-Bielefeld collaboration we obtained $\chi^2=20.9$, $T_{ch}=165.9$ MeV, and $\mu_b=20.9$ MeV. This provides further evidence \cite{NoronhaHostler:2007jf,longpaper,NoronhaHostler:2007fg,NoronhaHostler:2008ju} that Hagedorn states should be included in a description of hadronic matter near $T_c$. Since the chemical freeze-out temperature was found to increase from $160$ MeV to roughly $165$ MeV (RBC-Bielefeld) or $172$ MeV (BMW) when including Hagedorn states, this exemplifies the degree of uncertainty in extracting chemical freeze-out thermodynamical parameters by means of such thermal analyzes.

Furthermore, because Hagedorn states provide a bridge between the chemical freeze-out temperature and the critical temperature, we were able to make a qualified statement about which critical temperature region is more appropriate according to the quality of the thermal fits. We find that lower critical temperature regions are favored because more Hagedorn states are present close to the chemical freeze-out temperature and that a substantial number of Hagedorn states (i.e. a contribution of about 40$\%$ to the total particle numbers) are needed in order to provide the best fit to the hadron yield ratios.

A lower $\chi^2$ can be obtained for the higher critical temperature region when we double the degeneracy of the Hagedorn states, which would lead to a mismatch between our thermodynamic quantities and those computed on the lattice (recall that the parameters that define the exponential spectrum in this case are obtained by fitting the results of the RBC-Bielefeld collaboration at $\mu_b=0$). As we can see from Tab.\ \ref{tab:all} a change in the parameters even when they are doubled still gives a better fit than the thermal fits without Hagedorn states because a contribution of Hagedorn states as small as 4-11$\%$ still contribute enough to lower $\chi^2$. Therefore, this reconfirms the importance of including Hagedorn states in the hadron gas phase and, consequently, in the computation of thermal fits. Moreover, our results indicate that hadronization and chemical equilibration do not need to occur at the same temperature in order to explain RHIC data.

\section{Acknowledgments}

J.N. acknowledges support from US-DOE Nuclear Science Grant No. DE-FG02-93ER40764. This work was supported by the Helmholtz International Center
for FAIR within the framework of the LOEWE program (Landes-Offensive zur Entwicklung
Wissenschaftlich-¨okonomischer Exzellenz) launched by the State of Hesse.

%%%%%%%%%%%%%%%%%%%%%%%%%%%%%%%%%%%%%%%%%%%%%%%%%%%%%%%%%%%%%%%%%%%%%%%

%%%%%%%%%%%%%%%%%%%%%%%%%%%%%%%%%%%%%%%%%%%%%%%%%%%%%%%%%%%%%%%%%%%%%%%


\begin{thebibliography}{99}
\bibitem{Cheng:2007jq}
  M.~Cheng {\it et al.},
  %``The QCD Equation of State with almost Physical Quark Masses,''
  Phys.\ Rev.\  D {\bf 77}, 014511 (2008).
  %%CITATION = PHRVA,D77,014511;%%
\bibitem{Bazavov:2009zn}
  A.~Bazavov {\it et al.},
  %``Equation of state and QCD transition at finite temperature,''
  arXiv:0903.4379 [hep-lat].
\bibitem{zodor}
  Y.~Aokia {\it et al.},
  arXiv:hep-lat/0510084v2.

\bibitem{thermalmodels}
  %%CITATION = ARXIV:0903.4379;%%
  P.~Braun-Munzinger, K.~Redlich and J.~Stachel,
  %``Particle production in heavy ion collisions,''
  arXiv:nucl-th/0304013;
  P.~Braun-Munzinger, J.~Stachel, J.~P.~Wessels and N.~Xu,
  %``Thermal equilibration and expansion in nucleus-nucleus collisions at the
  %AGS,''
  Phys.\ Lett.\  B {\bf 344}, 43 (1995);Phys.\ Lett.\  B {\bf 365}, 1 (1996);
  J.~Cleymans, D.~Elliott, A.~Keranen and E.~Suhonen,
  %``Thermal model analysis of particle ratios at GSI Ni Ni experiments  using
  %exact strangeness conservation,''
  Phys.\ Rev.\  C {\bf 57}, 3319 (1998);
  J.~Cleymans, H.~Oeschler and K.~Redlich,
  %``Influence Of Impact Parameter On Thermal Description Of Relativistic Heavy
  %Ion Collisions At (1-2) A-Gev,''
  Phys.\ Rev.\  C {\bf 59}, 1663 (1999);
  R.~Averbeck, R.~Holzmann, V.~Metag and R.~S.~Simon,
  %``Neutral pions and eta mesons as probes of the hadronic fireball in  nucleus
  %nucleus collisions around 1-A-GeV,''
  Phys.\ Rev.\  C {\bf 67}, 024903 (2003);
  P.~Braun-Munzinger, I.~Heppe and J.~Stachel,
  %``Chemical equilibration in Pb + Pb collisions at the SPS,''
  Phys.\ Lett.\  B {\bf 465}, 15 (1999).
  J.~Cleymans, H.~Satz, E.~Suhonen and D.~W.~von Oertzen,
  %``STRANGENESS PRODUCTION IN HEAVY ION COLLISIONS AT FINITE BARYON NUMBER
  %DENSITY,''
  Phys.\ Lett.\  B {\bf 242}, 111 (1990);
  J.~Cleymans and H.~Satz,
  %``Thermal hadron production in high-energy heavy ion collisions,''
  Z.\ Phys.\  C {\bf 57}, 135 (1993);
  F.~Becattini, M.~Gazdzicki and J.~Sollfrank,
  %``On chemical equilibrium in nuclear collisions,''
  Eur.\ Phys.\ J.\  C {\bf 5}, 143 (1998);
  F.~Becattini, J.~Cleymans, A.~Keranen, E.~Suhonen and K.~Redlich,
  %``Features of particle multiplicities and strangeness production in  central
  %heavy ion collisions between 1.7-A-GeV/c and 158-A-GeV/c,''
  Phys.\ Rev.\  C {\bf 64}, 024901 (2001); G.~Torrieri and J.~Rafelski,
  %``Strange hadron resonances as a signature of freeze-out dynamics,''
  Phys.\ Lett.\  B {\bf 509}, 239 (2001); G.~Torrieri, S.~Steinke, W.~Broniowski, W.~Florkowski, J.~Letessier and J.~Rafelski,
  %``SHARE: Statistical hadronization with resonances,''
  Comput.\ Phys.\ Commun.\  {\bf 167}, 229 (2005);
  S.~Wheaton and J.~Cleymans,
  %``THERMUS: A thermal model package for ROOT,''
  Comput.\ Phys.\ Commun.\  {\bf 180}, 84 (2009);


A.~Kisiel, T.~Taluc, W.~Broniowski and W.~Florkowski,
  %``THERMINATOR: Thermal heavy-ion generator,''
  Comput.\ Phys.\ Commun.\  {\bf 174}, 669 (2006) .
\bibitem{StatModel}
  C.~Spieles, H.~Stoecker and C.~Greiner,
  %``Hadron production in relativistic nuclear collisions: Thermal hadron
  %source or hadronizing quark-gluon plasma?,''
  Eur.\ Phys.\ J.\  C {\bf 2}, 351 (1998)
\bibitem{Schenke:2003mj}
  B.~Schenke and C.~Greiner,
  %``Statistical description with anisotropic momentum distributions for  hadron
  %production in nucleus nucleus collisions,''
  J.\ Phys.\ G {\bf 30}, 597 (2004).

\bibitem{RHIC}
  P.~Braun-Munzinger, D.~Magestro, K.~Redlich and J.~Stachel,
  %``Hadron production in Au Au collisions at RHIC,''
  Phys.\ Lett.\  B {\bf 518}, 41 (2001);
  W.~Florkowski, W.~Broniowski and M.~Michalec,
  %``Thermal analysis of particle ratios and p(T) spectra at RHIC,''
  Acta Phys.\ Polon.\  B {\bf 33}, 761 (2002);
  W.~Broniowski and W.~Florkowski,
  %``Strange particle production at RHIC in a single-freeze-out model,''
  Phys.\ Rev.\  C {\bf 65}, 064905 (2002);
  M.~Kaneta and N.~Xu,
  %``Centrality dependence of chemical freeze-out in Au + Au collisions at
  %RHIC,''
  arXiv:nucl-th/0405068;
  J.~Adams {\it et al.}  [STAR Collaboration],
  %``Experimental and theoretical challenges in the search for the quark  gluon
  %plasma: The STAR collaboration's critical assessment of the  evidence from
  %RHIC collisions,''
  Nucl.\ Phys.\  A {\bf 757}, 102 (2005).

\bibitem{Andronic:2005yp}
  A.~Andronic, P.~Braun-Munzinger and J.~Stachel,
  %``Hadron production in central nucleus nucleus collisions at chemical
  %freeze-out,''
  Nucl.\ Phys.\  A {\bf 772}, 167 (2006).
  %%CITATION = NUPHA,A772,167;%%
%\cite{Manninen:2008mg}
\bibitem{Manninen:2008mg}
  J.~Manninen and F.~Becattini,
  %``Chemical freeze-out in ultra-relativistic heavy ion collisions at
  %sqrt(s)_NN = 130 and 200 GeV,''
  Phys.\ Rev.\  C {\bf 78}, 054901 (2008).
  %%CITATION = PHRVA,C78,054901;%%
\bibitem{freezeoutline}
  P.~Braun-Munzinger and J.~Stachel,
  %``Dynamics of ultra-relativistic nuclear collisions with heavy beams: An
  %experimental overview,''
  Nucl.\ Phys.\  A {\bf 638}, 3 (1998);
  J.~Cleymans and K.~Redlich,
  %``Unified description of freeze-out parameters in relativistic heavy ion
  %collisions,''
  Phys.\ Rev.\ Lett.\  {\bf 81}, 5284 (1998);Phys.\ Rev.\  C {\bf 60}, 054908 (1999);
  J.~Cleymans,
  %``Rapidity and energy dependence of thermal parameters,''
  J.\ Phys.\ G {\bf 35}, 044017 (2008);
  J.~Cleymans, R.~Sahoo, D.~K.~Srivastava and S.~Wheaton,
  %``Saturation of Transverse Energy per Charged Hadron and Freeze-Out Criteria
  %in Heavy-Ion Collisions,''
  Eur.\ Phys.\ J.\ ST {\bf 155}, 13 (2008)
  J.~Cleymans, R.~Sahoo, D.~P.~Mahapatra, D.~K.~Srivastava and S.~Wheaton,
  %``Transverse Energy per Charged Particle and Freeze-Out Criteria in Heavy-Ion
  %Collisions,''
  Phys.\ Lett.\  B {\bf 660}, 172 (2008);
  J.~Cleymans, H.~Oeschler, K.~Redlich and S.~Wheaton,
  %``Status of chemical freeze-out,''
  J.\ Phys.\ G {\bf 32}, S165 (2006).

\bibitem{LHC}
  N.~Armesto {\it et al.},
  %``Heavy Ion Collisions at the LHC - Last Call for Predictions,''
  J.\ Phys.\ G {\bf 35}, 054001 (2008).
  %%CITATION = JPHGB,G35,054001;%%


\bibitem{Rapp:2000gy}
  R.~Rapp and E.~V.~Shuryak,
  %``Resolving the antibaryon production puzzle in high-energy heavy-ion
  %collisions,''
  Phys.\ Rev.\ Lett.\  {\bf 86} (2001) 2980.
  %%CITATION = PRLTA,86,2980;%%
\bibitem{Greiner}
  C.~Greiner,
  %``Importance of multi-mesonic fusion processes on (strange) antibaryon
  %production,''
  AIP Conf.\ Proc.\  {\bf 644}, 337 (2003);
  Heavy Ion Phys.\  {\bf 14}, 149 (2001);
  %%CITATION = NUCL-TH 0011026;%%
  C.~Greiner and S.~Leupold,
  %``Antihyperon production in relativistic heavy ion collision,''
  J.\ Phys.\ G {\bf 27}, L95 (2001).
\bibitem{Koch:1986ud}
  P.~Koch, B.~Muller and J.~Rafelski,
  %``Strangeness In Relativistic Heavy Ion Collisions,''
  Phys.\ Rept.\  {\bf 142}, 167 (1986).
  %%CITATION = PRPLC,142,167;%%
%\cite{Stock:1999hm}
\bibitem{Kapusta}
  J.~I.~Kapusta and I.~Shovkovy,
  %``Thermal rates for baryon and anti-baryon production,''
  Phys.\ Rev.\ C {\bf 68} (2003) 014901;
  %%CITATION = NUCL-TH 0209075;%%
  J.~I.~Kapusta,
  %``Thermal rates for baryon and anti-baryon production,''
  J.\ Phys.\ G {\bf 30} (2004) S351.
  %%CITATION = JPHGB,G30,S351;%%
%\cite{Huovinen:2003sa}
\bibitem{Huovinen:2003sa}
  P.~Huovinen and J.~I.~Kapusta,
  % ``Rate equation network for baryon production in high energy nuclear
  %collisions,''
  Phys.\ Rev.\ C {\bf 69} (2004) 014902.
  %%CITATION = NUCL-TH 0310051;%%
\bibitem{Stock:1999hm}
  R.~Stock,
  %``The parton to hadron phase transition observed in Pb + Pb collisions at
  %158-GeV per nucleon,''
  Phys.\ Lett.\  B {\bf 456}, 277 (1999).
  %%CITATION = NUCL-TH/0703050;%%
\bibitem{Heinz:2006ur}
  U.~Heinz and G.~Kestin,
  %``Universal chemical freeze-out as a phase transition signature,''
  arXiv:nucl-th/0612105.
  %%CITATION = NUCL-TH/0612105;%%
\bibitem{BSW}
  P.~Braun-Munzinger, J.~Stachel and C.~Wetterich,
  %``Chemical freeze-out and the QCD phase transition temperature,''
  Phys.\ Lett.\ B {\bf 596} (2004) 61.
  %%CITATION = NUCL-TH 0311005;%%
\bibitem{Hagedorn:1968jf}
  R.~Hagedorn,
  % ``Statistical thermodynamics of strong interactions at high energies. 3.
  %Heavy-pair (quark) production rates,''
  Nuovo Cim.\ Suppl.\  {\bf 6}  311 (1968);
  Nuovo Cim.\ Suppl.\  {\bf 3}, 147 (1965).
\bibitem{Greiner:2004vm}
  C.~Greiner {\it et al.}
  %``Chemical equilibration due to heavy Hagedorn states,''
  J.\ Phys.\ G {\bf 31}, S725 (2005).
  %%CITATION = HEP-PH 0412095;%%
\bibitem{NoronhaHostler:2007jf}
  J.~Noronha-Hostler, C.~Greiner and I.~A.~Shovkovy,
  %``Fast Equilibration of Hadrons in an Expanding Fireball,''
  Phys.\ Rev.\ Lett.\  {\bf 100}, 252301 (2008).
  %%CITATION = PRLTA,100,252301;%%
\bibitem{longpaper}
  J.~Noronha-Hostler, C.~Greiner and I.~A.~Shovkovy,
  to appear.
\bibitem{NoronhaHostler:2007fg}
  J.~Noronha-Hostler, C.~Greiner and I.~A.~Shovkovy,
  %``Chemical equilibration at the Hagedorn temperature,''
  Eur.\ Phys.\ J.\ ST {\bf 155}, 61 (2008);
  arXiv:nucl-th/0703079.
  %%CITATION = NUCL-TH/0703079;%%
\bibitem{NoronhaHostler:2008ju}
  J.~Noronha-Hostler, J.~Noronha and C.~Greiner,
  %``Transport Coefficients of Hadronic Matter near $T_c$,''
  arXiv:0811.1571 [nucl-th].
  %%CITATION = ARXIV:0811.1571;%%
\bibitem{KSS}
  P.~Kovtun, D.~T.~Son and A.~O.~Starinets,
  %``Viscosity in strongly interacting quantum field theories from black hole
  %physics,''
  Phys.\ Rev.\ Lett.\  {\bf 94}, 111601 (2005).
  %%CITATION = PRLTA,94,111601;%%
%\cite{NoronhaHostler:2008ju}

\bibitem{Kharzeev:2007wb}
  D.~Kharzeev and K.~Tuchin,
  %``Bulk viscosity of QCD matter near the critical temperature,''
  JHEP {\bf 0809}, 093 (2008); F.~Karsch, D.~Kharzeev and K.~Tuchin,
  %``Universal properties of bulk viscosity near the QCD phase transition,''
  Phys.\ Lett.\  B {\bf 663}, 217 (2008).


%\cite{Broniowski:2004yh}
\bibitem{Broniowski:2004yh}
  W.~Broniowski, W.~Florkowski and L.~Y.~Glozman,
  %``Update of the Hagedorn mass spectrum,''
  Phys.\ Rev.\  D {\bf 70}, 117503 (2004).
  %%CITATION = PHRVA,D70,117503;%%

%\cite{Cohen:2009wq}
\bibitem{Cohen:2009wq}
  T.~D.~Cohen,
  %``QCD and the Hagedorn spectrum,''
  arXiv:0901.0494 [hep-th].
  %%CITATION = ARXIV:0901.0494;%%

%\cite{Bugaev:2008nu}
\bibitem{Bugaev:2008nu}
  K.~A.~Bugaev, V.~K.~Petrov and G.~M.~Zinovjev,
  %``Why Don't We See the Hagedorn Mass Spectrum in the Experiments?,''
  arXiv:0801.4869 [hep-ph];
  K.~A.~Bugaev, V.~K.~Petrov and G.~M.~Zinovjev,
  %``Fresh look at the Hagedorn mass spectrum as seen in the experiments,''
  arXiv:0812.2189 [nucl-th].
%\cite{Moretto:2006zz}
\bibitem{Moretto:2006zz}
  L.~G.~Moretto, L.~Phair, K.~A.~Bugaev and J.~B.~Elliott,
  %``The Hagedorn Thermostat,''
  PoS C {\bf POD2006} (2006) 037;
  L.~G.~Moretto, K.~A.~Bugaev, J.~B.~Elliott and L.~Phair,
  %``Can a Hagedorn system have a temperature other than $T_C$ or can a
  %thermostat have a temperature other than its own?,''
  arXiv:nucl-th/0601010;
  arXiv:hep-ph/0511180.
\bibitem{Pal:2005rb}
  S.~Pal and P.~Danielewicz,
  %``Hadron production from resonance decay in relativistic collisions,''
  Phys.\ Lett.\  B {\bf 627}, 55 (2005).
  %%CITATION = PHLTA,B627,55;%%

\bibitem{Kapusta:1982qd}
  J.~I.~Kapusta and K.~A.~Olive,
  %``Thermodynamics Of Hadrons: Delimiting The Temperature,''
  Nucl.\ Phys.\  A {\bf 408}, 478 (1983).
  %%CITATION = NUPHA,A408,478;%%
%\cite{Rischke:1991ke}
\bibitem{Rischke:1991ke}
  D.~H.~Rischke, M.~I.~Gorenstein, H.~Stoecker and W.~Greiner,
  %``Excluded Volume Effect For The Nuclear Matter Equation Of State,''
  Z.\ Phys.\  C {\bf 51}, 485 (1991).
  %%CITATION = ZEPYA,C51,485;%%

\bibitem{Eidelman:2004wy}
  S.~Eidelman {\it et al.}
  Phys.\ Lett.\  B {\bf 592}, 1 (2004).

\bibitem{Liu}
  F.~M.~Liu, K.~Werner and J.~Aichelin,
  %``Comparison of microcanonical and canonical hadronization,''
  Phys.\ Rev.\ C {\bf 68} (2003) 024905;
  %%CITATION = HEP-PH 0304174;%%
  F.~M.~Liu, et. al.,
   %``A micro-canonical description of hadron production in proton proton
  %collisions,''
  J.\ Phys.\ G {\bf 30} (2004) S589;
  Phys.\ Rev.\ C {\bf 69} (2004) 054002.
  %%CITATION = EPHJA,C38,225;%%
\bibitem{STAR}
 O.~Y.~Barannikova  [STAR Collaboration],
  %``Probing collision dynamics at RHIC,''
  arXiv:nucl-ex/0403014;
  J.~Adams {\it et al.}  [STAR Collaboration],
  %``Experimental and theoretical challenges in the search for the quark  gluon
  %plasma: The STAR collaboration's critical assessment of the  evidence from
  %RHIC collisions,''
  Nucl.\ Phys.\  A {\bf 757}, 102 (2005).
\bibitem{PHENIX}
   S.~S.~Adler {\it et al.}  [PHENIX Collaboration],
  %``Identified charged particle spectra and yields in Au + Au collisions at
  %s(NN)**(1/2) = 200-GeV,''
  Phys.\ Rev.\  C {\bf 69}, 034909 (2004).

\end{thebibliography}
\end{document}